# Hadron locations on the rectangle

B. F. Riley

AMEC NNC Limited,
601 Faraday Street, Birchwood Park,
Warrington WA3 6GN, UK

bernard.riley@amecnnc.com

**Abstract**

Hadrons are located at the orthogonal intersections of 4-branes that wrap 1-cycles of a rectangular compact space. The 4-branes are arranged in a regular network. We reveal the conspicuous locations on the rectangle of the lightest mesons, unflavoured and of each flavour, of the lightest meson singlet states with hidden flavour of each kind, and of the lightest baryons of each flavour. The Higgs field vacuum expectation value is also associated with a conspicuous position on the rectangle, at the orthogonal intersection of 4-branes.

# 1 Introduction

In the Randall and Sundrum RS1 model [1], the Standard Model particles and forces are confined to a 3-brane, the 'weak' brane, at a fixed point of the $S^1/Z_2$ orbifold. The Planck brane is located at the other fixed point of the orbifold. The two branes are separated by a slice of AdS spacetime, of curvature $k$. The natural scale on the weak brane is suppressed from Planck-scale by a warp factor exp(-$k$y), where y is the coordinate of the extra dimension. In the RS2 model [2], there is only one brane, at y = 0 in an infinite dimension, upon which the Standard Model particles and forces are confined. Lykken and Randall have combined the results of RS1 and RS2 to address the hierarchy problem in an infinite dimension [3]. In this model, gravity is localized on the Planck brane and we live on a brane in the fifth dimension. In the *infinitely large new dimensions* model of Arkani-Hamed, Dimopoulos, Dvali and Kaloper [4], our world is located at the orthogonal intersection of (2+n)-branes in $AdS_{4+n}$ spacetime. In our six-dimensional model, massive particles are located at the orthogonal intersections of 4-branes that wrap 1-cycles of a rectangular compact space [5, 6]. The 4-branes are arranged in a regular network.

First, we review the model and then show that the lightest mesons, unflavoured and of each flavour, the lightest meson singlet states with hidden flavour of each kind, and the lightest $J^P = \frac{1}{2}^+$ baryons of each flavour are located at conspicuous positions on the rectangle, at the intersections of 4-branes. We also show that the weak gauge bosons $W^\pm$ and $Z^0$ are associated with a similarly conspicuous position, as is the Higgs field vacuum expectation value.

Values of particle mass have been taken from the Particle Listings 2006 of the Particle Data Group [7].

# 2 The model

The bulk spacetime of the model is AdS and the higher dimensional Planck scale is equal in size to that of the four-dimensional Planck Mass $M_P$ (1.221 x $10^{19}$ GeV). We relate four-dimensional mass parameters $m$ to positions $y$ in an extra dimension through the equation

$$m = e^{-ky} M_P. \qquad (1)$$

A lattice of points with spacing $d/k$ in a semi-infinite extra dimension corresponds to a geometric sequence of mass-energy scales that descends from the Planck Mass $M_P$ with common ratio exp(-$d$). Particles of all types are found to occupy the levels and sublevels, described below, of two such geometric sequences: Sequence 1 and Sequence 2, of common ratio $2/\pi$ and $1/\pi$, respectively. The mass $m_i$ of the $i^{th}$ level of Sequence 1 is given by

$$m_i = (\pi/2)^{-i} M_P, \qquad (1)$$

where $i \geq 0$. The mass $m_j$ of the $j^{th}$ level of Sequence 2 is given by

$$m_j = \pi^{-j} M_P, \qquad (2)$$

where $j \geq 0$. Principal mass levels in Sequence 1 and Sequence 2 are of integer $i$ and $j$, respectively. Higher order mass levels (sublevels) are of fractional $i$ and $j$. For example, $1^{st}$ order levels in Sequence 1 are of half-integer $i$, while $2^{nd}$ order levels are of quarter-integer $i$. The dimensionless variables $i$ and $j$ measure the corresponding distance along the two extra dimensions, from the Planck region $(i, j) = (0, 0)$, in units of $(1/k)\ln(\pi/2)$ and $(1/k)\ln\pi$, respectively. At the principal lattice points in the two extra dimensions, $i$ and $j$ are of integer value. Higher order lattice points divide the space between the principal ($0^{th}$ order) lattice points into $2^n$ intervals of equal length, where n is the order.

The two lattices we have identified lie in two directions, of coordinates $y$ and $z$, in a noncompact two-dimensional extra space. 4-branes extending through the lattice points of one dimension, in a direction parallel to the second dimension, intersect 4-branes extending through the lattice points of the second dimension, in a direction parallel to the first dimension, and partition the AdS spacetime. 4-branes of $n^{th}$ order extend through lattice points of $n^{th}$ order. The major scales of physics correspond to the positions of low order 4-brane intersections $(i, j)$ lying on the line $y = z$ [4]. We have provided evidence that the noncompact extra space of the model is the infinite covering space of a rectangular compact space with sides of length $(1/k)\ln(\pi/2)$ and $(1/k)\ln\pi$ [5]. Quarks, hadrons and the weak gauge bosons are all associated with the orthogonal intersections, on the line $y = z$ in the covering space, of 4-branes wrapping 1-cycles of the rectangle.

## 3  Hadron locations on the rectangle

On the rectangle, meson and baryon singlet states are located upon the intersections of 4-branes. Hadrons comprising an isospin doublet, or multiplets in which the particles are of one of two masses, are centred on such intersections. That is, the geometric mean of the two masses corresponds to the position of a 4-brane intersection in the covering space. Since the massive particles of our world are evidently constrained to lie on the line y = z in the covering space [4], they will not be located precisely upon the intersections of 4-branes on the rectangle. Instead, they tend to be located close to low order intersections, each particle taking up a location precisely upon one 4-brane of low order, and upon an orthogonal 4-brane of higher order [5].

The locations on the rectangle of the lightest mesons, unflavoured and of each flavour, are shown in Figure 1. Each isospin multiplet ($\pi^\pm$ - $\pi^0$, $K^\pm$ - $K^0$, $D^\pm$ - $D^0$ and $B^\pm$ - $B^0$) is located upon a 4-brane of 4th or lower order, and upon an orthogonal 4-brane of higher order. Also shown in Figure 1 are the locations of the predominantly $s\bar{s}$ state $\phi$ and the lightest $c\bar{c}$ meson $\eta_c$, upon low order 4-brane intersections. The lightest meson singlet state $\eta$ and the lightest $b\bar{b}$ meson Y are located upon higher order 4-brane intersections, as shown in Figure 2, which represents a portion of the rectangle.

The lightest $J^P = \frac{1}{2}^+$ baryons of each flavour are associated with 4-branes of 0th and 1st order lying parallel to the $z$ direction, as shown in Figure 3. The uds baryons $\Lambda$ and $\Sigma^0$ form a doublet of sorts, centred on a 0th order 4-brane. The mass difference of these two baryons equals the mass of a 0th order level in Sequence 1, within the small measurement uncertainty. The udc baryon $\Lambda_c^+$ and the udb baron $\Lambda_b^0$ are located upon a 1st order 4-brane. The ssc baryon $\Omega_c^0$ and the exotic baryon $\theta^+$ are located upon 4-branes of 3rd order. The proton – neutron isospin doublet (not shown) is centred precisely upon a 4-brane of 7th order.

## 4  The Higgs field VEV

As a postscript to this paper, we show in Figure 4 the location on the rectangle of the $W^\pm$ - $Z^0$ doublet [5], the location a 115 GeV Higgs boson would occupy, and the position corresponding to the Higgs field vacuum expectation value (246.22 GeV). Each of these locations is found at the orthogonal intersection of 4-branes.

**References**

1. L. J. Randall and R. Sundrum, *A large mass hierarchy from a small extra dimension*, Phys. Rev. Lett. **83**, 3370 (1999), hep-ph/9905221
2. L. J. Randall and R. Sundrum, *An alternative to compactification*, Phys. Rev. Lett. **83**, 4690 (1999), hep-th/9906064
3. J. Lykken and L. J. Randall, *The shape of gravity*, JHEP 0006 (2000) 014, hep-th/9908076 N.
4. Arkani-Hamed, S. Dimopoulos, G. Dvali and N. Kaloper, *Infinitely large new dimensions*, Phys. Rev. Lett. **84**, 586 (2000), hep-th/9907209
5. B. F. Riley, *Mass hierarchies from two extra dimensions*, physics/0603014
6. B. F. Riley, *The intersecting brane world*, physics/0607152
7. W.–M. Yao et al., J. Phys. G**33**, 1 (2006)

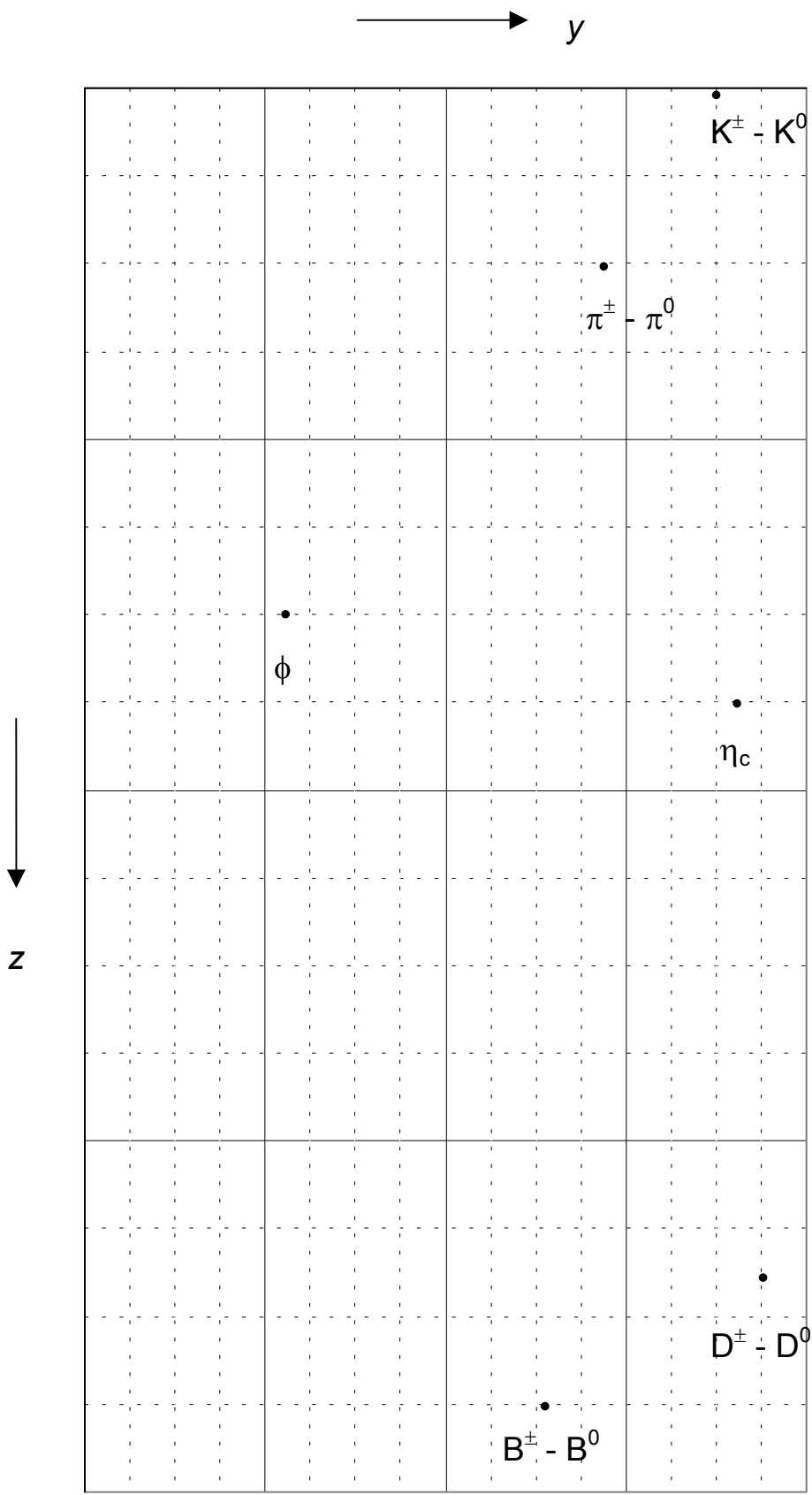

Figure 1: Meson locations on the rectangle

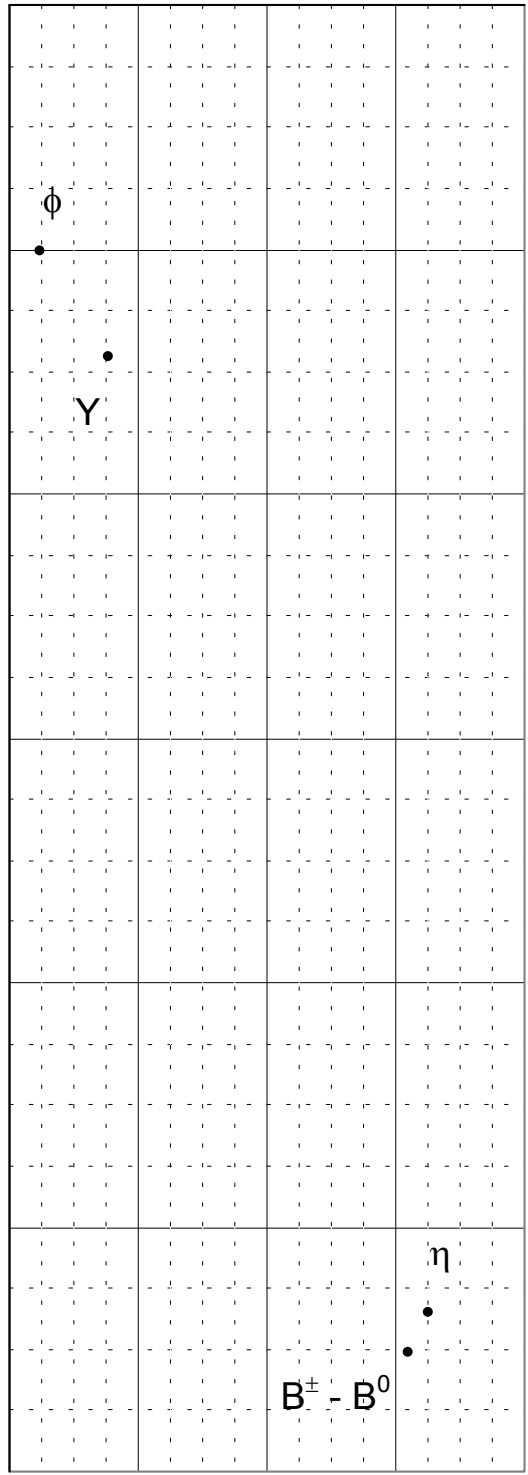

Figure 2: Meson locations on a portion of the rectangle

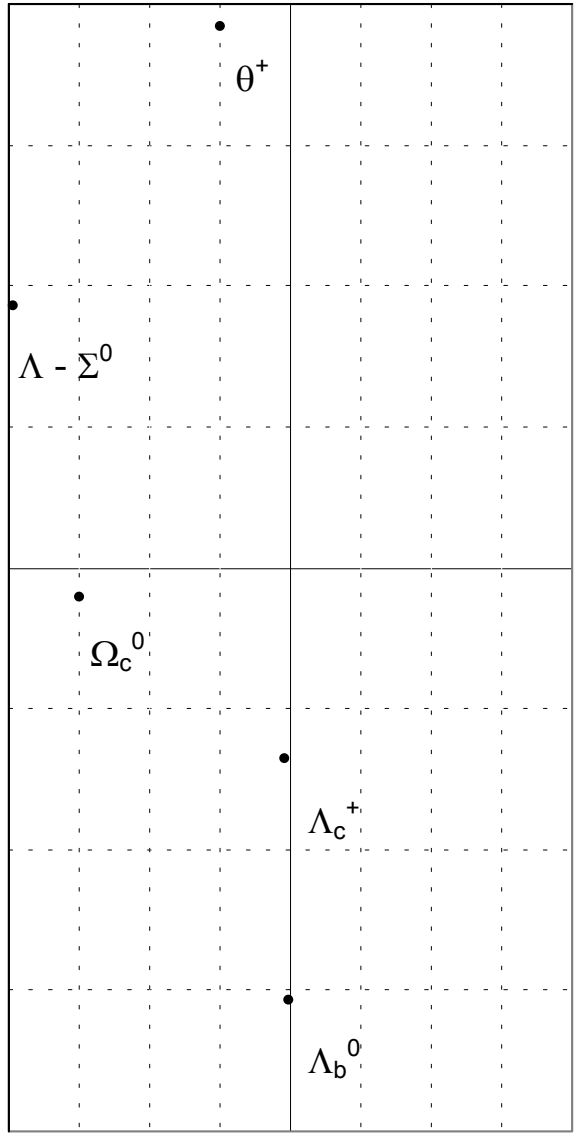

Figure 3: Baryon locations on the rectangle

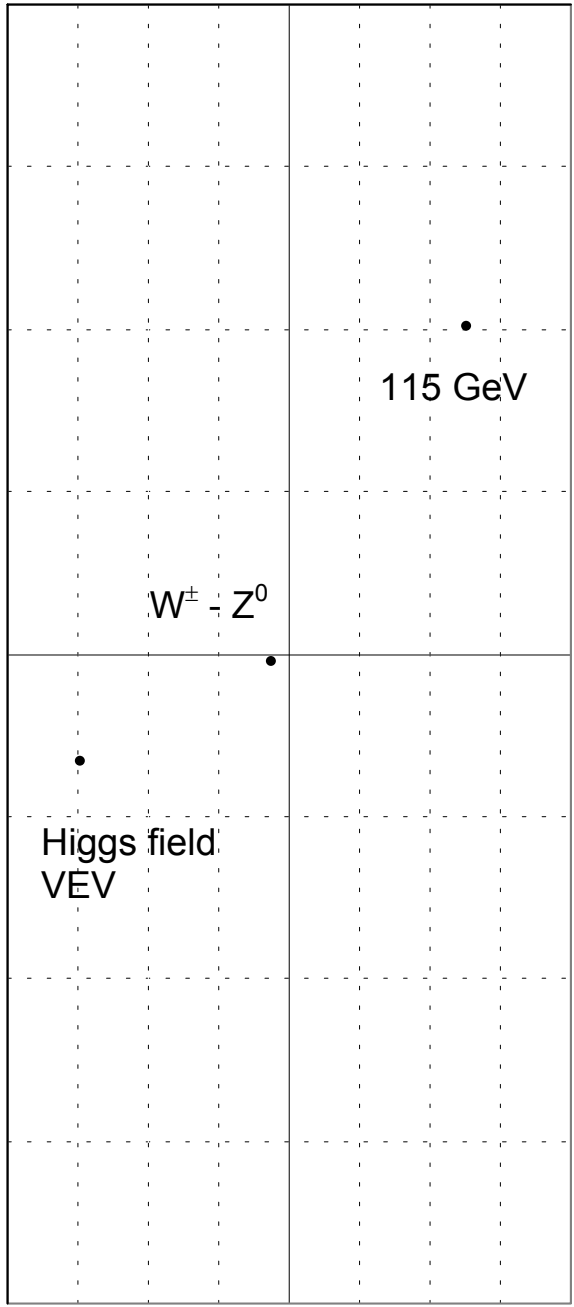

Figure 4: Other locations on the rectangle